\documentclass[final,5p,times,twocolumn]{elsarticle}
\usepackage{graphicx}
\usepackage{amssymb}
\usepackage{lineno}
\usepackage{color}

\journal{ }

\begin{document}

\newcommand{\red}{\textcolor{red}}
\newcommand{\blue}{\textcolor{blue}}
\newcommand{\etc}{\textit{etc.}}
\newcommand{\ie}{\textit{i.e.\,}}
\newcommand{\eg}{\textit{e.g.\,}}
\newcommand{\etal}{\textit{et al.\,}}
\newcommand{\roha}{Rohacell\textsuperscript{\textregistered} }
\newcommand{\rohacomma}{Rohacell\textsuperscript{\textregistered}, }
\newcommand{\Tedlar}{Tedlar\textsuperscript{\textregistered} }
\newcommand{\Tedlarcomma}{Tedlar\textsuperscript{\textregistered}, }

\begin{frontmatter}

\title{The Design and Performance of a Scintillating-Fibre Tracker for the Cosmic-ray Muon Tomography of Legacy Nuclear Waste Containers}

\author[Glasgow]{A.\,Clarkson}
\author[Glasgow]{D.\,J.\,Hamilton}
\author[Glasgow]{M.\,Hoek\fnref{Curr}}
\author[Glasgow]{D.\,G.\,Ireland}
\author[NNL]{J.\,R.\,Johnstone}
\author[Glasgow]{R.\,Kaiser}
\author[Glasgow]{T.\,Keri}
\author[Glasgow]{S.\,Lumsden}
\author[Glasgow]{D.\,F.\,Mahon}
\author[Glasgow]{B.\,McKinnon}
\author[Glasgow]{M.\,Murray}
\author[Glasgow]{S.\,Nutbeam-Tuffs}
\author[NNL]{C.\,Shearer}
\author[NNL]{C.\,Staines}
\author[Glasgow]{G.\,Yang}
\author[NNL]{C.\,Zimmerman}

\address[Glasgow]{SUPA, School of Physics \& Astronomy, University of Glasgow, Kelvin Building, University Avenue, Glasgow, G12 8QQ, Scotland, UK}
\address[NNL]{National Nuclear Laboratory, Central Laboratory, Sellafield, Seascale, Cumbria, CA20 1PG, England, UK}
\fntext[Curr]{Current affiliation:  Johannes Gutenberg-Universit\"{a}t, Mainz}

\begin{abstract}
Tomographic imaging techniques using the Coulomb scattering of cosmic-ray muons are increasingly being exploited for the non-destructive assay of shielded containers in a wide range of applications.  One such application is the characterisation of legacy nuclear waste materials stored within industrial containers.  The design, assembly and performance of a prototype muon tomography system developed for this purpose are detailed in this work.  This muon tracker comprises four detection modules, each containing orthogonal layers of Saint-Gobain BCF-10 2\,mm-pitch plastic scintillating fibres.  Identification of the two struck fibres per module allows the reconstruction of a space point, and subsequently, the incoming and Coulomb-scattered muon trajectories.  These allow the container content, with respect to the atomic number $Z$ of the scattering material, to be determined through reconstruction of the scattering location and magnitude.  On each detection layer, the light emitted by the fibre is detected by a single Hamamatsu H8500 MAPMT with two fibres coupled to each pixel via dedicated pairing schemes developed to ensure the identification of the struck fibre.  The PMT signals are read out to QDCs and interpreted via custom data acquisition and analysis software.

The design and assembly of the detector system are detailed and presented alongside results from performance studies with data collected after construction.  These results reveal high stability during extended collection periods with detection efficiencies in the region of 80\% per layer.  Minor misalignments of millimetre order have been identified and corrected in software.  A first image reconstructed from a test configuration of materials has been obtained using software based on the Maximum Likelihood Expectation Maximisation algorithm.  The results highlight the high spatial resolution provided by the detector system.  Clear discrimination between the low, medium and high-$Z$ materials assayed is also observed.

\end{abstract}


\end{frontmatter}


\section{Introduction}
When high-energy cosmic rays bombard the Earth, muons are produced as part of particle showers within the upper atmosphere.  These highly-penetrating particles are observed at sea level with a flux of approximately one per square centimetre per minute and momenta of several GeV\,c$^{-1}$.  As charged particles, they interact with matter primarily through ionising interactions with atomic electrons and via Coulomb scattering from nuclei.  Both of these mechanisms have been exploited in recent years in the field of Muon Tomography (MT) to probe the internal composition of shielded structures which cannot be probed using convential forms of imaging radiation \eg X-rays.   Since E.\,P. George measured the thickness of the ice burden above the Guthega-Munyang tunnel in Australia in the 1950s~\cite{George1955}  and L.\,W. Alvarez conducted his search for hidden chambers in the Second Pyramid of Chephren in Egypt~\cite{Alvarez1970} a decade later, there has been a wealth of wide-ranging applications using cosmic-ray muons for imaging purposes, such as in the field of volcanology~\cite{Tanaka2005,Ambrosi2011} and nuclear contraband detection for national security~\cite{Borozdin2003a,Gnanvo08}.

The seminal work outlined by Borozdin \etal in Ref.~\cite{Borozdin2003a} revealed the potential to locate and characterise materials within shielded containers using the Coulomb scattering of cosmic-ray muons.  This approach relies on precision reconstruction of the initial and scattered muon trajectories to determine the scattering location within the container and the scattering density, denoted $\lambda$.  This scattering density is known to exhibit an inherent dependence on the atomic number $Z$ of the material~\cite{Schultz04} \ie larger scattering angles are typically observed for objects with larger $Z$ values. 

This work presents the design and fabrication processes of a small prototype MT system for use in the identification and characterisation of legacy nuclear waste materials stored in highly-engineered industrial waste containment structures.  For this purpose, a detector with high spatial resolution was required.  It was essential that the design of the system, and the materials and fabrication processes used, be scaleable to allow the future construction of an industrially deployable detector system.  The requirement of industrial deployability mandated that the detection medium be radiation-hard, robust and capable of performing to a high degree of stability over prolonged periods of time for operation in the typical high radiation, high stress industrial environment found within a nuclear waste processing plant.

A modular design based on plastic scintillating fibres was chosen to satisfy these criteria.  The individual components and modular construction are described in Section~\ref{sec:Components} with the assembly process outlined in Section~\ref{sec:Assembly}.  The readout electronics, data acquisition (DAQ) and complex fibre multiplexing requirements are described in Section~\ref{sec:Mapping}.   Results showing the performance of the constructed, commissioned system are presented in Section~\ref{sec:Commissioning} and first results from the image reconstruction of a test configuration of low, medium and high-$Z$ materials are presented in Section~\ref{sec:Images}. 

\section{Tracker Module Design \& Components}\label{sec:Components}
Each detector module comprised orthogonal detection planes of plastic scintillating fibres which were supported and positioned on a low-density rigid foam and aluminium structure.  The two planes of fibres were bonded within this structure and further supported at the extremities by plastic polymer distribution blocks.  Fibres were coupled to photon detectors (here, multi-anode photomultiplier tubes) housed within lightproof boxes.  The active area of each module was encased within a lightproof vinyl film.  The design of the individual detector modules and all components and materials used in the fabrication process are detailed in the following section. 

\subsection{Scintillating Fibres \& Hamamatsu H8500 PMTs}
Prior to the design and fabrication processes, dedicated Geant4~\cite{geant4} simulation studies were performed to assess the fibre pitch required for this system with regards to the anticipated light output and reconstructed image resolution~\cite{SimPaper}.  A design based on 2\,mm-pitch plastic scintillating fibres was chosen.  The fibres used in the production of this muon tracker were Saint-Gobain\footnote{http://www.saint-gobain.com} BCF-10 round fibres comprising a polystyrene-based core (97\% by cross-sectional width) with polymethylmethacrylate (PMMA) optical cladding (3\%).  The light emission output of this formulation, which peaked at 432\,nm, provided excellent overlap with the sensitivity of the chosen photon detector, the Hamamatsu\footnote{http://www.hamamatsu.com} H8500 MAPMT.

Feasibility studies (outlined in Section~\ref{sec:Mapping}) revealed the potential to reduce the required number of readout channels by 50\% through the coupling of two scintillating fibres to a single pixel on the 64-pixel segmented anode of the PMT.  A single detection layer comprising 128 fibres coupled to a single PMT was therefore chosen for the final prototype design. 

The design for the complete detector system comprised four modules, two situated above, and two below the volume under interrogation \ie the assay volume.  With this modular design and the identification of a space point in each module, the initial and scattered muon trajectories could be reconstructed allowing the determination of the scattering position and magnitude required for image reconstruction.  Within each module, two orthogonal detection planes, comprising a single layer of 128 fibres, yielded an active area of 256\,mm~x~256\,mm.   To allow for minimal transmission losses and manageable strain as a result of the fibre bending required for PMT coupling, fibres of 860\,mm length were used.  In total, for eight detection layers, 1024 fibres and 8 PMTs were required.

\begin{figure}[t] 
\centering 
\includegraphics[width=\columnwidth,keepaspectratio]{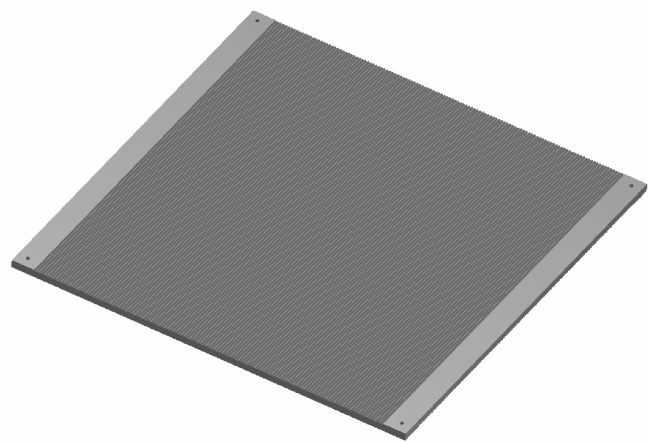}
\caption{A CAD schematic of a \roha support sheet etched with 128 \textit{V} grooves (central region).  The four holes on the smooth, non-grooved edges were used to secure the locating pins on the aluminium base plate described in Section~\ref{sub:AluminiumBaseplate}. }
\label{fig:Rohacell}
\end{figure}

\subsection{\roha Support Sheets}\label{sub:roha}
For each fibre layer, accurate fibre positioning and support was provided by a low-$Z$, precision-machined structure of \roha sheeting\footnote{manufactured by Evonik Industries (www.rohacell.com)},  a closed-cell rigid polymethacrylimide foam.  This provided sufficiently high tensile strength in comparatively small thicknesses to support the individual fibre layers whilst providing only a negligible Coulomb scattering effect on transient muons.   A layered configuration of \roha support sheets of varying thicknesses and dimensions was fabricated to support the orthogonal layers of fibres.  The assembly of this structure is described in detail in Section~\ref{sub:Module}.   

Shown in Figure~\ref{fig:Rohacell} and measuring 300\,mm in length, the central square sheet that supported the fibres was machined with shallow, parallel \textit{V} grooves to position each of the 128 fibres per layer.  Grooves were cut in the x (y) direction on the top (bottom) side of the support sheet.  On this central sheet, narrow non-grooved regions were located at the sides parallel to the fibre direction (shown in Figure~\ref{fig:Rohacell}) to accommodate the holes which allowed the sheet to be fixed onto locating pins on the aluminium base plate (see Section~\ref{sub:AluminiumBaseplate}).  Each sheet of \roha had locating holes in each corner to fix the sheet to these locating pins to ensure accurate alignment and uniformity across all four modules.  

\begin{figure}[t] 
\centering 
\includegraphics[width=\columnwidth,keepaspectratio]{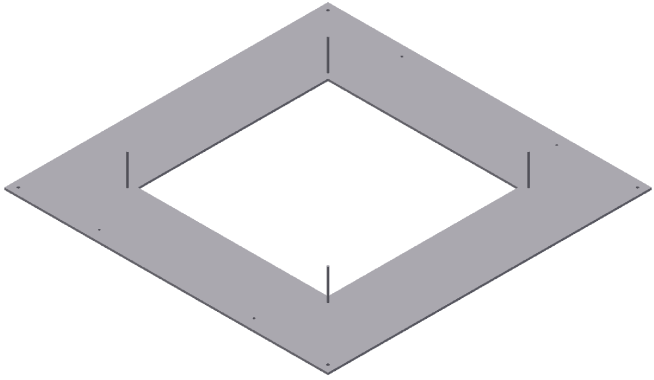}
\caption{Schematic of the aluminium base plate showing the removed section of the active area, the stainless steel locating pins, and the holes for the fibre distribution blocks and aluminium profile frame.  The plate dimensions are described in the text.}
\label{fig:Al-plate}
\end{figure}
\begin{figure}[b] 
\centering 
\includegraphics[width=\columnwidth,keepaspectratio]{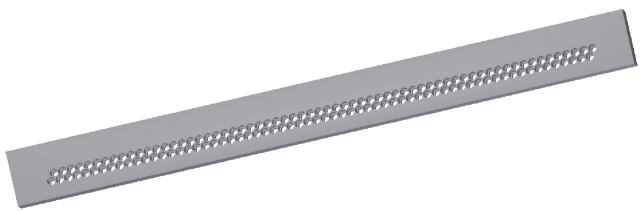}
\caption{The rectangular fibre distribution POM block showing the two staggered rows of holes.  These blocks measured 300\,mm in length and supported the fibres at the extremes of the aluminium base plate.  Not shown are the screws used to fix the block to the base plate.}
\label{fig:FibreDistBlock}
\end{figure}

\subsection{Aluminium Base Plate}\label{sub:AluminiumBaseplate}
A 3\,mm-thick flat sheet of aluminium was used to provide a robust support base for the multiple layers of \roha and scintillating fibres which formed the active area of each module.  This square sheet of aluminium, shown in Figure~\ref{fig:Al-plate}, measured 460\,mm with a central square region of 270\,mm removed to minimise the material contributing towards Coulomb scattering within the active volume.  Four stainless-steel locating pins were fixed at each corner of the inner hole in the base plate. These were used to secure the sheets of \roha and to provide fixed reference points which ensured precision alignment was maintained both internally (\ie with the fibres within the corresponding module) and externally (\ie with the other three modules).  At the sides of the aluminium plate, two locating holes were bored.  These holes (the innermost holes on Figure~\ref{fig:Al-plate}) allowed fibre distribution blocks to be located on each axis and screwed into place.  Each side of the base plate had two further holes bored at the outer corners.  These ensured the aluminium sheet could be fixed to an external aluminium profile stand.

\subsection{Fibre Distribution Blocks}
To provide extra support to the fibres, custom-built distribution blocks were fabricated from polyoxymethylene (POM) plastic. These also maintained the uniform distribution of fibres at the extremes of the base plate and, more importantly, at the pixelated surface of the PMT.  This was chosen over other potentially abrasive candidate materials \eg aluminium,  to prevent causing damage to the fibres.  The detector benefitted from the low cost, opacity and precision machining capability of this polymer.  

The first set of distribution blocks were rectangular POM pieces shown in Figure~\ref{fig:FibreDistBlock}.  These were located at the extreme of the base plate and in the same plane as the fibre layers.  They contained two rows of 64 holes and measured 300\,mm in length.  The rows were both offset slightly, one above and one below, from the level of the fibres to provide support and stability to the fibres and to prevent unwanted strain causing them to lift off from the \roha surface after bonding.  This also ensured that a constant tension was maintained across all fibres.   The fibre distribution blocks were fixed in place on the aluminium base by screws through the holes in the plate.  

The holes which supported the fibres were large enough to allow the application of black nylon tubing to the fibres for lightproofing.  This tubing covered the length of each individual fibre outwith the area covered by the base plate and mitigated the risk of external sources of light incident on the fibre from producing signals within the PMT.  This tubing, and additional lightproofing measures, are outlined in Section~\ref{sub:lighttight}.

\begin{figure}[t] 
\centering 
\includegraphics[width=\columnwidth,keepaspectratio]{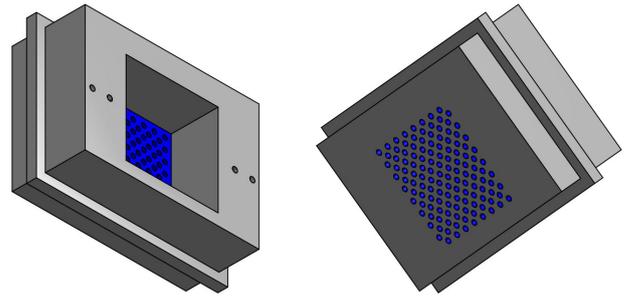}
\caption{A CAD schematic of the dual-purpose (fibre distribution and PMT holder) POM block.  The 128 holes which hold the fibres in position at the PMT surface are observed within the PMT cavity (left) and the front end facing the active area (right).  The raised areas along the edges are used to secure the holder to the aluminium profile support stand.  The four holes on the surface (left) allow the attachment of the printed circuit board (PCB) for readout (see Section~\ref{sec:Mapping}).}
\label{fig:PMTholder}
\end{figure}

\begin{figure}[t] 
\centering 
\includegraphics[width=\columnwidth,keepaspectratio]{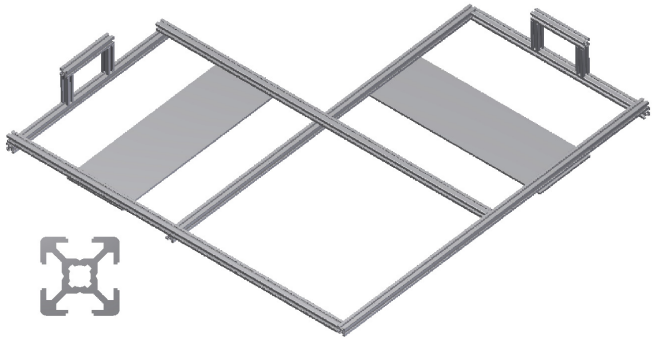}
\caption{Schematic of the aluminium profile frame for each detector module (main image).  Shown are active area (central region) and the orthogonal arms which support the x and y planes of fibres.  On each arm, aluminium sheeting provides further support for the fibres close to the PMTs.  The specially-constructed housings for the PMT holders are shown at the extremes of the arms.  The cross-section of the aluminium profile is also shown (bottom left).}
\label{fig:frame}
\end{figure}

The second POM block shown in Figure~\ref{fig:PMTholder} measured 110\,mm\,x\,82\,mm\,x\,42\,mm.  It served to hold the PMT in place in an inner square cavity (of equal dimensions to the PMT) at the rear side, whilst at the front, it provided precision contact between the pixelated surface of the PMT and each coupled fibre pair. This latter characteristic was achieved via an array of 128 holes, positioned such that two fibres were in optical contact with a single pixel.  Once inserted into the holes of the PMT holder, the fibres protruded a short distance due to a series of stainless steel collars which were bonded to the end of the fibres (see Section~\ref{sub:fibreassembly}).  These prevented the fibres extending further into the PMT holder and thus ensured that the small lengths of fibre that protruded uniformly contacted the PMT surface.  This process also ensured that all the fibres exerted the same pressure on the surface of the PMT.  A thin silicone gel protective pad, which had negligible effect on the transmission of light, was placed between the fibres and PMT to mitigate potential damage to the surface of the PMT by the pressure of the fibres.  Four threaded holes were positioned on the rear surface to allow a Printed Circuit Board (PCB) to be connected for signal readout purposes.  Surrounding the edge of the PMT holder, a raised border allowed the holder to sit firmly within a specially-constructed housing on the aluminium profile frame, shown in Figure~\ref{fig:frame}.

\subsection{Aluminium Profile Module Frame}
All components outlined in the previous subsections were housed in an aluminium profile frame (with 20\,mm square cross-sectional area) shown schematically in Figure~\ref{fig:frame}.  Aluminium profile was chosen because of its lightweight structure and strength.  This was constructed to support the base plate (onto which it was screwed), \roha and orthogonal layers of fibres, in addition to preventing unwanted strain on the fibres in the non-active regions which could result in fibre breakages and/or loss of contact with the PMT surface.  To minimise these risks, aluminium sheets were attached to the frame structure to provide further support to the fibres in this region.  At the end of these two arms, three small aluminium profile struts were assembled to fix the PMT holders firmly in position.  These struts were designed to accommodate the raised borders of the holders.  The design of the module frame benefitted from the capability of aluminium profile to attach to an external support frame at any position and orientation along its length.  This allowed complete freedom for future experimental studies on alignment and module positioning. 

\subsection{Lightproof PMT Housing}
\begin{figure}[t] 
\centering 
\includegraphics[width=\columnwidth,keepaspectratio]{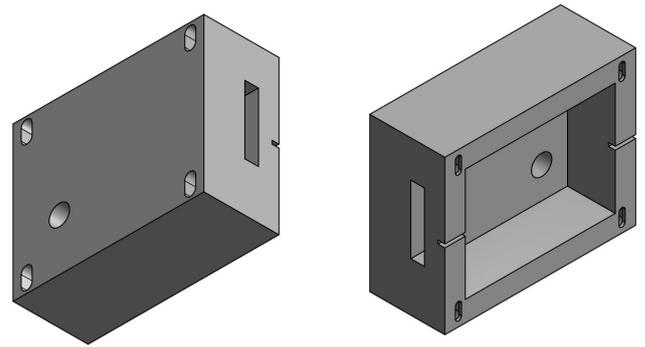}
\caption{CAD schematic of the POM box constructed to enclose the PMT, PCB, and their readout and power cables.  On both views, the four access holes which secured the box to the aluminium profile, and the circular hole for the PMT signal cable are shown.  At the edges, a small rectangular slit allowed the PMT high-voltage supply cable to be connected.  All holes are covered with black electrical tape once all cables have been connected.  The larger rectangular slit at the side wall accommodated the quartet of ribbon cables used to read out the pixel signals.}
\label{fig:PMTbox}
\end{figure}

Custom-made PCBs were printed to couple to the H8500 PMT connector pins and allow the connection of 16-channel ribbon cables to facilitate the readout of signals from the pixels.  The readout system will be described in detail in Section~\ref{sec:Mapping}.  These PCBs were tightly-screwed to the PMT holder via the holes on either side.  This helped provide uniform contact between the fibres and the face of the PMT.  A final, rectangular POM box measuring 140\,mm\,x\,113\,mm\,x\,55\,mm was fabricated and bolted to the aluminium profile to fully encase and lightproof the PMT and the associated voltage supply cables, the PCB and readout signal cables.   This POM box is shown in Figure~\ref{fig:PMTbox}.  This was bolted to the aluminium profile of the module via four oblique holes in the corners.  A rectangular slit at one side of the box allowed the four ribbon signal cables per PMT to be connected.  A small groove at each end of the box provided access for the high voltage cable and ensured it was not damaged during assembly.  A circular hole was drilled out at the back of the PMT cover to allow the signal cable to be inserted.

\section{Assembly of the Prototype Detector System}\label{sec:Assembly}
This section will detail the preparatory and assembly processes associated with the fabrication of the detector modules, and the lightproofing measures taken.

\begin{figure*}[t] 
\centering 
\includegraphics[width=2\columnwidth,keepaspectratio]{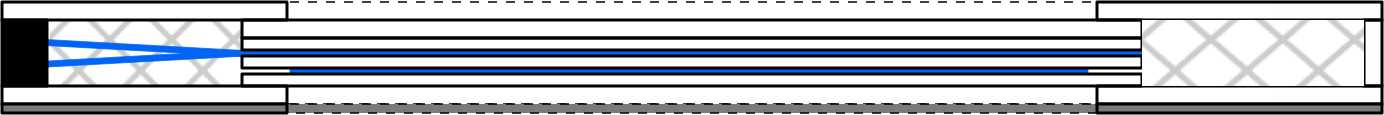}
\caption{Diagram showing a cross-section through the centre (in the $y$ direction) of the layered structure of a completed module.  Shown are the Rohacell\textsuperscript{\textregistered} support sheets (white), scintillating fibres (blue), aluminium (grey) layers and fibre distribution block (black).  Top-down the Rohacell\textsuperscript{\textregistered} layers measure thicknesses of 6\,mm, 6\,mm, 4\,mm, 6\,mm (with grooves on the topside and underside in alternate planes), 4\,mm and 6\,mm.  The outermost \roha layers, and the 3\,mm-thick aluminium baseplate have removed sections in the centre (indicated by the dashed lines) of the active area to minimise the amount of material in this region.  At the left (right) hand side, the fibre distribution block (\roha skirting) is shown.}
\label{fig:modulelayers}
\end{figure*}
\subsection{Preparation of the Scintillating Fibres}\label{sub:fibreassembly}
Prior to assembly, both ends of the scintillating fibres were polished to optimise light transmission and to provide good optical contact with the pixelated surface of the PMTs.  A three-stage polishing process was undertaken using a series of lapping films of decreasing coarseness.

A vital component of the assembly process was to maintain a uniform contact and pressure distribution between the fibres and the PMT.  This was achieved using a series of metal tubes (or collars) fixed to one end of the fibres at precise positions.  At this end of each fibre, a 30\,mm-long stainless steel collar was bonded with optical epoxy ensuring that a small length of fibre protruded from the collar.  Once the epoxy and collar had been applied, the epoxy was left to cure.  After curing, a further lapping stage removed this protruding length of fibre and, with it, any epoxy covering the polished end.  

A smaller 5\,mm-long stainless steel collar was bonded onto the larger collar using a UV-curing adhesive.  Removable caps were placed on the fibres during this process to ensure that the small collars were all attached at a pre-set position along the larger collar.  This ensured that every collared fibre protruded the same distance through the PMT holder, providing a uniform pressure on the PMT surface.  This process of attaching the smaller collars was performed after the fibres had been bonded to the \roha sheets and covered in black nylon tubing, which are detailed in the following sections. 

\begin{figure*}[t] 
\centering 
\includegraphics[width=\columnwidth,keepaspectratio]{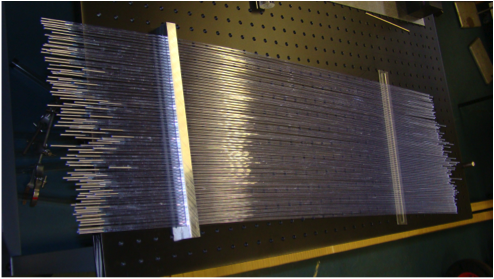}
\includegraphics[width=\columnwidth,keepaspectratio]{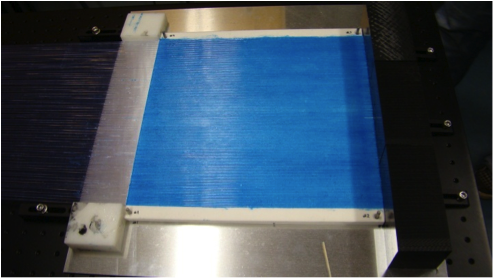}
\caption{The polished and collared fibres displayed within the fibre distribution block pre-assembly (left) and a single layer of scintillating fibres bonded to the grooved sheet of \roha (right).   Here, the blue, epoxy-coated region corresponds to the active area.  On the right hand side of this image, the black POM blocks used to ensure alignment of the fibres with the layers of \roha are shown temporarily clamped to the aluminium base plate on the breadboard surface.}
\label{fig:Fibres}
\end{figure*}

\subsection{Construction of the Detector Modules}\label{sub:Module}
Figure~\ref{fig:modulelayers} illustrates the final layered structure of scintillating fibres, \roha and aluminium which comprised each detector module.  The scintillating fibres were first bonded to the central, grooved \roha sheet using the same optical epoxy used for the bonding of the large collars. 

The grooved sheet of \roha was first attached to the aluminium base plate via the four locating pins.  This base plate was clamped to a breadboard table with flat blocks of POM abutting the end of the \roha sheet, as shown in Figure~\ref{fig:Fibres}.  This ensured that, when bonded, the open end of the fibres rested level with the edge of their support sheet.  The repellent properties of POM ensured that the epoxy did not bond to these blocks during this process.  Epoxy was then applied to the top side of the sheet before fibres (these would become the bottom $y$ plane) were carefully positioned into the grooves.  A flat, 4\,mm-thick sheet of \roha of otherwise equal dimensions to the grooved central sheet, was then placed over the locating pins and bonded on top of the layer of fibres.  This was weighted evenly to ensure the fibres cured in their desired positions within the grooves.   With this layer of fibres set in place, the configuration was then overturned to allow the top x plane of fibres to be bonded.   This was placed onto the base plate once again, over a 6\,mm-thick piece of \roha of equal dimensions as the aluminium plate.  To minimise the material in the active volume, the central square region of this piece, measuring 270\,mm, was also removed.

For the x plane of fibres, the same gluing procedure was performed with another 4\,mm-thick sheet of \roha bonded and weighted on top during the curing process.   A 6\,mm-thick flat piece of \rohacomma the same dimensions as the other central pieces, was secured on top of the support pins.  This sheet was necessary to ensure a level height with the POM distribution blocks screwed to the edge of the base plate.  The fibres were then fed through these blocks with care taken to avoid damage.  A final sheet with the central region removed was pinned to the structure to fully enclose the fibres.   This was secured tightly with washers over the four locating pins.  Once the layered configuration had been fabricated, a \roha skirting, shown in Figure~\ref{fig:modulelayers}, was pinned around the open edges of the module to seal off the active area and to provide a smooth-edged support for the lightproof covering described in the following subsection.  This whole assembly was then secured in the aluminium profile module frame using purpose-built bolts placed in the troughs of profile (shown previously in Figure~\ref{fig:frame}) which were then bolted to the holes in the base plate.   Each module was then prepared for final lightproofing measures.   

\subsection{Lightproofing Measures}\label{sub:lighttight}
Prior to the optical connection of the fibres to the PMTs, the exposed regions outwith the encased active area, and the areas covered by the porous \rohacomma required lightproofing to prevent light from entering the active area and producing noise signals during data collection.  To achieve this, lengths of black nylon tubing were applied over the collared ends (prior to the bonding of the smaller collar) forming a tight seal against the large aluminium collar, while leaving sufficient space for the small collar to be bonded.  At the opposite end, the tubing was positioned tightly inside the holes of the fibre distribution block providing a firm seal.  When all the fibres were covered in the tubing, each fibre was labelled sequentially to allow the fibre multiplexing schemes outlined in Section~\ref{sec:Mapping} to be implemented.  

\begin{figure*}[t] 
\centering 
\includegraphics[width=\columnwidth,keepaspectratio]{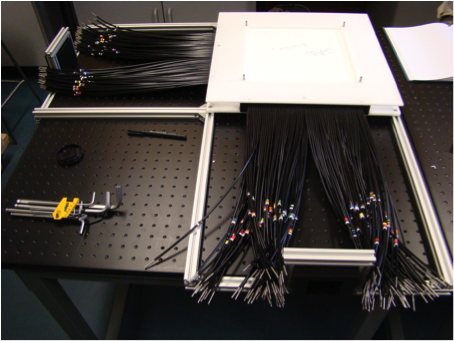}
\includegraphics[width=\columnwidth,keepaspectratio]{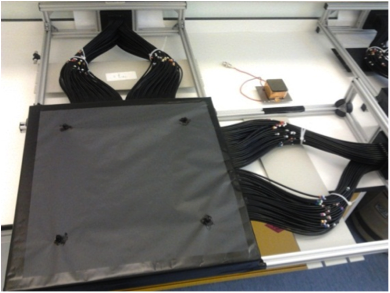}
\caption{A detector module (left) showing the orthogonal layers of fibres and \roha attached to the aluminium profile frame prior to the application of \Tedlar.  The PMT holders are observed in the bottom and top-left of the image.  Black nylon tubing, coloured labels used for identification, and collars are shown around the lengths of the fibres outwith the active area.   A fully lightproof detector module (right) within its aluminium profile frame.  The black \Tedlar film and nylon tubing protective coverings are shown.  Also shown are the electrical taping and locating pin coverings described in the text.  Here, the fibres in both planes are shown connected to the PMT via the array of holes in the POM blocks.  Not shown are the black, lightproof POM boxes which house the PMTs and PCBs.}
\label{fig:Tedlar}
\end{figure*}

The \roha structure around the active area was covered in sheets of \Tedlarcomma a radiation-durable polyvinyl fluoride film\footnote{manufactured by DuPont\textsuperscript\texttrademark (www.dupont.com)} with excellent light absorption properties, which ensured the fibres were not exposed to any external light.  Two layers of this film were wrapped around the central area of the module with four holes pierced in the film to allow the locating pins to protrude through.  Once in place, black electrical tape was applied around all the edges to hold the \Tedlar in position and a black foam seal was placed over the locating pins and taped in place to restrict any light penetrating into the active area.  The fully lightproof, and PMT-connected, detector module is shown in Figure~\ref{fig:Tedlar}. 

\subsection{Vertical Support Stand}
With the construction of the four detector modules completed, and lightproof testing successfully performed, the modules were assembled in an aluminium profile vertical support stand.  The modules were precisely aligned, to make certain that all four were parallel, and positioned at predetermined, optimum positions in the vertical $z$ direction.   These positions were extensively studied using detailed Geant4 simulations of the detector geometry to optimise the tracked muon flux and reconstructed image resolution.  A maximum separation of 900\,mm between the outermost modules was imposed to negate any destabilising effects introduced by extending to longer lengths of aluminium profile for the vertical stand.  Limiting this to 900\,mm ensured the stability and alignment of the system.  The separations between the top and bottom module pairs were fixed at 250\,mm leaving a 400\,mm spacing in the assay volume to accommodate test objects for future image reconstruction.

The individual modules were attached to the support stand via a series of clamps around the central region. These ensured firm and stable attachment to the vertical stand, and allowed the modules to be connected at any position in the vertical direction if required.   The clamps were designed with two holes on the outside and one on the raised, middle section.  This middle hole provided the ability for the clamp to be bolted onto the vertical struts of the external detector stand, while the outside holes bolted onto each module frame.  The four clamps were positioned at the active area of each module with two on the outside of the frame and two on the inner side.  

\begin{figure}[t] 
\centering 
\includegraphics[width=\columnwidth,keepaspectratio]{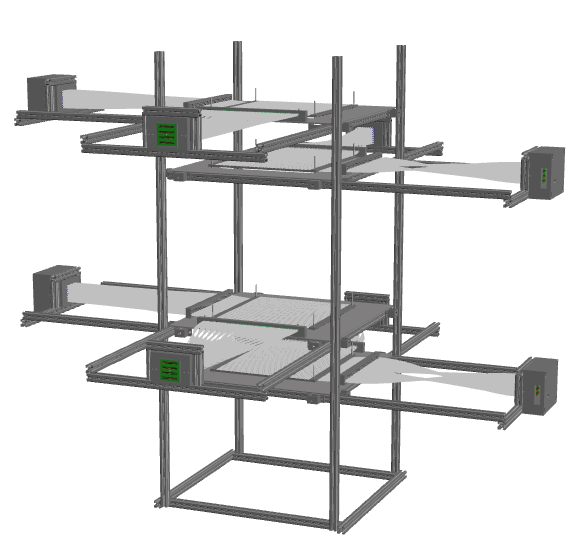}
\caption{A CAD model of the prototype MT system.   Shown are the four tracking modules consisting of two orthogonal layer of scintillating fibres each connected to a PMT.  These are housed in the POM distribution blocks and holders described in the text.  The four modules are assembled in an aluminium profile support frame. }
\label{fig:GMTdetector}
\end{figure}

A model of the detector, fully assembled in the vertical support structure, is shown in Figure~\ref{fig:GMTdetector}.  The square base of the vertical support stand was firmly secured to a breadboard surface to provide additional stability.  Alignment studies, presented in Section~\ref{sub:Alignment}, revealed the high degree of precision resulting from the various support measures taken during the assembly of this detector system.


\section{Data Acquisition \& Signal Readout Systems}\label{sec:Mapping}
The Hamamatsu PMTs offer one negative-voltage read out channel per pixel and a unified signal from the second-to-last dynode (Dynode-12 or Dy12) in the amplification chain with positive polarity that is sensitive to a signal in any pixel. The Dy12 signals were used in this system to trigger the read out and recording of the PMT signal. In order to reduce noise, a coincidence requirement was imposed; data were only recorded when three separate PMTs recorded signals in any pixel.

The system made use of a NIM electronics crate housing coincidence gates, dual gate generators, discriminators and fan-in/fan-out units. Data collection was performed via a custom C++ Data AcQuisition program (DAQ) running on a commodity laptop. This was connected via USB to a CAEN\footnote{http://www.caentechnologies.com} VME crate and the pixel signals were fed isomorphically into the signal inputs of CAEN charge to digital converter units (QDCs). The VME crate also housed a scaler unit that was used to monitor signal peaks above a pre-set threshold voltage.

The positive polarity Dy12 signals were fed into the NIM electronics crate, inverted, amplified and shaped before being passed to a set of coincidence units that produced the final physics trigger.

When the trigger condition was satisfied and the QDCs had recorded the analog signal, the DAQ software transferred the data from the QDC registers over the VME bus. The data were zipped and written to disk before the next event was recorded.   This potential delay had no significant impact on the detection efficiency due to the relatively slow event rate. Meanwhile, the scalers monitored the number of signals received from each PMT (each Dy12 signal was fed to an individual scalar channel), the number of read out gates generated by the NIM hardware, the number of trigger gates generated by the DAQ software and the number of events that were read out and recorded. For each event, the data and scaler signals were packaged together with a timestamp and status information on the QDC modules. The information from the scaler was used for both hardware and software debugging purposes; any light leaks are reflected in the number of signals generated by the PMTs while inefficiencies would be found if the number of physics triggers and the number of final triggers diverged. 

During normal data collection, events were recorded in runs of 500.  At the beginning of every data run, pedestal runs were recorded to monitor the amount of non-signal charge in the QDCs. As explained in Section 5, these pedestal files were used to identify any fluctuations in the electronics performance and were used during data quality monitoring to correct the corresponding data signals from the system prior to PMT gain correction.

\begin{figure}[t] 
\centering 
\includegraphics[width=0.8\columnwidth,keepaspectratio]{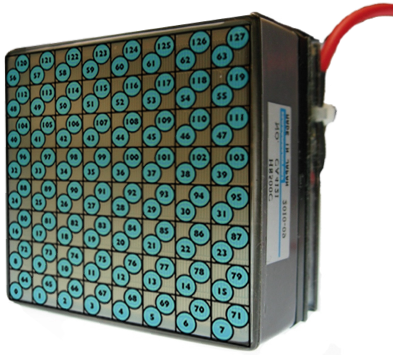}
\caption{Schematic of one of the fibre multiplexing schemes showing how each fibre pair couples to the 8\,x\,8 pixelated-anode structure of the surface of the Hamamatsu H8500 MAPMT.  Here, the fibre size is exaggerated in comparison with the pixel for illustration purposes.}
\label{fig:Mapping}
\end{figure}
\subsection{Fibre Coupling to PMTs}\label{sub:demultiplexing}
In order to minimise the number of read-out channels required for the experimental setup, the signals from all 128 scintillating fibres per detection layer were read out by a single Hamamatsu H8500 MAPMT.  Here the anode is segmented into a 64-pixel square array, with each pixel measuring 5.8\,mm in width.   Two fibres were coupled to each pixel via dedicated pairing (or multiplexing) schemes, an example of which is shown in Figure~\ref{fig:Mapping}.  

Simulation studies performed using a Geant4 simulation of the detector geometry prior to the construction of the prototype system showed that these multiplexing schemes allowed the correct identification of the eight struck fibres per event, thus maintaining the high spatial resolution provided by the 2\,mm pitch.  This was achieved using a likelihood-based demultiplexing algorithm which relied on a small scattering angle assumption and the use of four unique multiplexing schemes, one in each of the four detection layers in the top and bottom pairs of modules.  These had the effect of exaggerating the detected incident angles of the two muon trajectories for incorrectly-identified fibres such that the two partial tracks do not combine, or if so, are significantly less likely to have a scattering angle below a given threshold.

From these simulation studies, the correct eight fibres struck per event were successfully identified in over 98\% of events.   Further studies have shown this small misidentification of fibres which remains in the imaged data to have a negligible detrimental effect on the images reconstructed.

\section{Commissioning \& Performance Studies}\label{sec:Commissioning}
Once construction of the prototype detector system was completed and all DAQ systems had been tested, data were collected for commissioning and alignment purposes.  For the studies performed throughout this work, excluding the PMT gain normalisation and signal transmission studies (the experimental setups for these tests are described in Sections~\ref{sub:gain} and~\ref{sub:fibretest} respectively), the three-fold DAQ trigger described previously was implemented.  To recap, the top and bottommost layers were included in coincidence with one of the innermost layers to ensure confidence that the trigger originated from a single muon passing through the complete detector acceptance.  

For studies performed using the complete system (studies in Section~\ref{sub:QDC} onwards), dedicated pedestal-only and data runs were recorded at regular intervals during data collection.  The identification of the struck fibre was made by first requiring events in which the five layers outwith the trigger conditions registered a signal, in at least one channel, with a QDC signal 3$\sigma$ (from a Gaussian fit to the pedestal-only data) above the pedestal mean.   The pixel (or pixel cluster) coupled to the struck fibre was identified by selecting the highest pedestal-subtracted, gain-normalised QDC signal per PMT.  This yielded two potential fibres per detection layer as a consequence of the fibre multiplexing schemes.  The eight struck fibres per event, and therefore the four space points, were reconstructed via the likelihood-based demultiplexing algorithm described previously.

\subsection{PMT Characterisation \& Gain Normalisation}\label{sub:gain}
Prior to data collection on the full prototype system, each of the eight Hamamatsu H8500 MAPMTs to be used required detailed characterisation to assess the anticipated effects from cross-talk, clustering and local gain variations across the pixelated surface.  

Results from independent, detailed studies on the cross-talk and clustering characteristics of the H8500 PMT in Ref.~\cite{Montgomery12} by Montgomery \etal revealed these effects to be small.   Observations made during data collection consolidated these findings, revealing the influence of cross-talk on the surface of the PMT to be small in relation to the desired signal.  Typical cluster sizes of 2 to 3 pixels, and mean cluster multiplicities of around 1.5 per PMT were observed.   Although coupled on diagonally-opposite corners of the pixel and not the centre, the fibres (with collars) only occupied 23\% of the active area of the pixel.  This ensured that the majority of the signal was deposited within a single, distinct pixel.  The effect of cross-talk resulting from light transmission to neighbouring fibres was also investigated and found to have a negligible influence.  There was also no significant contribution arising from any dispersive effects introduced by the protective silicone pad applied to the face of each PMT.  As a consequence of this small cross-talk effect, the pixel corresponding to the QDC channel with the highest gain-normalised, pedestal-subtracted signal per PMT was considered as being coupled to the struck fibre.  Despite this, in approximately 5\% of events, there existed a secondary (or in even fewer cases, a tertiary) cluster with a larger integrated signal than the cluster containing the pixel with the highest individual signal.  Assignment of the corresponding fibres in such instances (in the vast majority of cases this equated to a single layer) manifested clear discrepancies in alignment data which were not observed when the pixel with the highest signal was selected.  These clusters were subsequently attributed to noise within the system.  It was concluded from these observations, that only the pixel with the highest signal per PMT should be considered for selection.  This was the case for all studies presented in this work. 

Negligence of any local gain variation effects across the PMT surface could, if their extent was large enough, result in the misidentification of the struck fibres through the selection of the PMT pixel with the highest signal.  To minimise the risk of this occurring, low and high resolution scans of each PMT were performed to determine the relative variations in gain across their surfaces.   These were performed using a laser tuned to the expected light output level of the scintillating fibres via a combination of filters.  Scans were undertaken at the operating voltages of the PMTs, with the low resolution scans stepping from the centre of each PMT pixel to the next, and the high resolution studies employing a step size of 1\,mm.   The efficiency and uniformity of the Dynode-12 signal were also studied during these laser scans to assess their potential use within the experimental trigger condition.

\begin{figure}[t] 
\centering 
\includegraphics[width=\columnwidth,keepaspectratio]{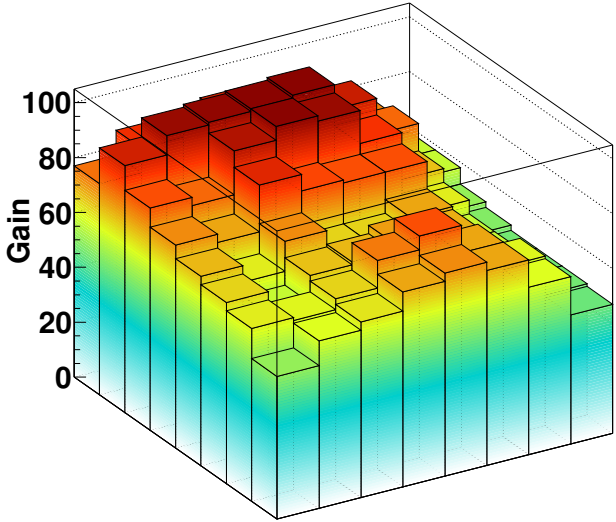}
\caption{Relative gain maps for one of the PMTs used in the constructed prototype system.  Shown are the gain values obtained via laser scans for the 64 PMT pixels relative to the highest pixel which is assigned a value of 100.  The small variations in gain across the pixelated surface of each PMT can give rise to ambiguities in the selection of the highest absolute signal (and hence the struck fibre) if each signal is not gain-normalised.}
\label{fig:gain}
\end{figure}

Results from the low resolution laser scans of one of the PMTs used in the prototype system are shown in Figure~\ref{fig:gain}.   These show the relative pixel gains normalised to the highest gain per PMT.   In rare, extreme cases, pixels are shown to drop to 40\% of the maximum gain of the PMT, with uniformities (defined here as the ratio of the maximum pixel gain to the mean gain) ranging from 1.2 to 1.6 across the eight PMTs.  With only a single PMT per detection layer on which to identify the hit pixel, it was not necessary to extend this localised normalisation across all eight PMTs.   
\begin{figure}[t] 
\centering 
\includegraphics[width=\columnwidth,keepaspectratio]{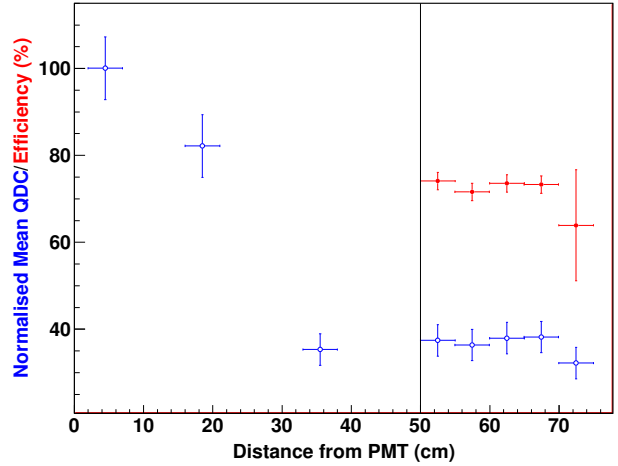}
\caption{Results from the light transmission tests on the 860\,mm-long scintillating fibres showing the mean normalised QDC signal (open blue circles) and detection efficiency (filled red circles) dependencies on the muon interaction position along the fibre relative to the PMT.  The mean QDC values have been normalised to a value of 100 at the PMT surface. Results show a large decrease in signal in the non-active area of the fibres which then remains stable across the active area (to the right of the dashed line) at a value of approximately 35\% of the signal observed at the PMT.  The detection efficiency remains constant across the active area.  Due to the bending of the fibres (and their irregular spread within this region) it was not possible to obtain accurate efficiency values outwith the active area.}
\label{fig:fibretest}
\end{figure}

\subsection{Fibre Signal Transmission Losses}\label{sub:fibretest}
The light transmission through the fibres was studied using a single detector module prior to assembly in the vertical support stand.  Small scintillator paddles, covering an area of approximately 50\,mm\,x\,50\,mm, were placed above and below the active area to provide an external muon trigger which removed any internal thresholds from biasing the efficiency measurement.   The entire length of the active area was scanned with selected positions along the remaining length of fibres to assess the extent of the anticipated degradation.  All measurements are presented normalised to the value (this is set to 100) obtained at the face of the PMT.

These studies confirmed the expected decrease in signal strength at increasing distances from the PMT due to the light transmission properties of the fibre, loss of light through the open end of the fibre, and the bending of the fibres at the PMT.   Figure~\ref{fig:fibretest} reveals the extent of this degradation in the observed signal in the region of this bending.  The signal remains relatively constant across the active area of the module at a value of approximately 35\% that of the signal observed at the PMT face.   The corresponding detection efficiencies (defined in Section~\ref{sub:eff}) also remain constant at values in the region of 75\% except in the final efficiency measurement.  It was subsequently discovered that the scintillator paddles did not fully cover the active area and as such, a larger uncertainty was attributed to account for this.  All results were obtained using the same 3$\sigma$ selection criteria for the signal size above pedestal, and confirmed that the degraded signals observed were still sufficiently large enough to avoid failing this requirement.

\begin{figure}[t] 
\centering 
\includegraphics[width=\columnwidth,keepaspectratio]{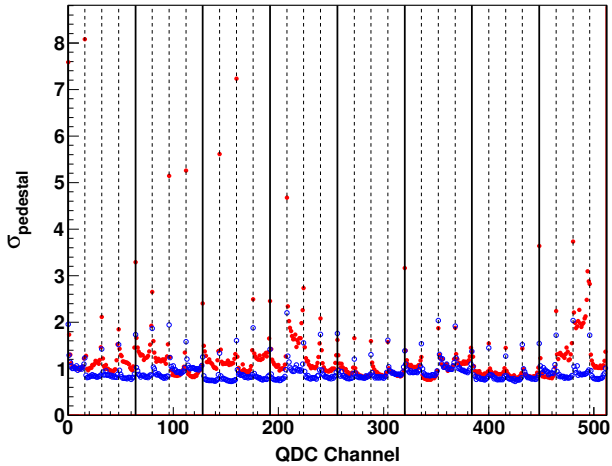}
\caption{Comparison between the pedestal $\sigma$ values observed from pedestal-only data for all 512 QDC channels prior to (red filled circles) and after (blue open circles) optimisation of the detector and readout setup.  In particular, channels in the QDC channel ranges [463, 478] and [479, 494] corresponded to two 16-channel ribbon cables which were later found to be severely kinked.  This induced noise on the corresponding channels.  In both sets of data, the characteristic parabolic distribution in $\sigma$ values is observed across the cable.  Here, the bold (dashed) lines separate regions belonging to the same detector layer (readout cable).}
\label{fig:sigmas}
\end{figure}

\subsection{QDC Signal Studies}\label{sub:QDC}
Throughout the commissioning process of the detector system, minor adjustments and improvements were made to the setup in preparation for image reconstruction studies.  Dedicated pedestal-only runs were recorded at regular intervals to allow stability assessment.  The widths of the pedestal distribution in each QDC channel, characterised by the $\sigma$ (alternatively, $\sigma_{\mathrm{pedestal}}$ in Figure~\ref{fig:sigmas}) from an applied Gaussian fit, were extracted.  Across all, the mean $\sigma$ was observed to be less than one QDC channel.  However, in channels read out by connectors at the edges of the ribbon cables, characteristically-broader pedestals were found with $\sigma$ values in the region of two channels.  This is an understood effect relating to the placement of the grounding pins.  The $\sigma$ values for each of the 512 read out channels are shown in Figure~\ref{fig:sigmas} for the optimised setup in comparison with initial pedestal data collected using what later transpired to be damaged ribbon cables which introduced noise into the system.  

\subsection{Detection Efficiency}\label{sub:eff}
In each of the five fibre layers that did not form part of the trigger scheme, the detection efficiencies were determined as the percentage of recorded events in which the layer registered at least one QDC signal more than 3$\sigma$ (from a Gaussian fit to the pedestal-only data) above the pedestal mean.    

\begin{figure}[t] 
\centering 
\includegraphics[width=\columnwidth,keepaspectratio]{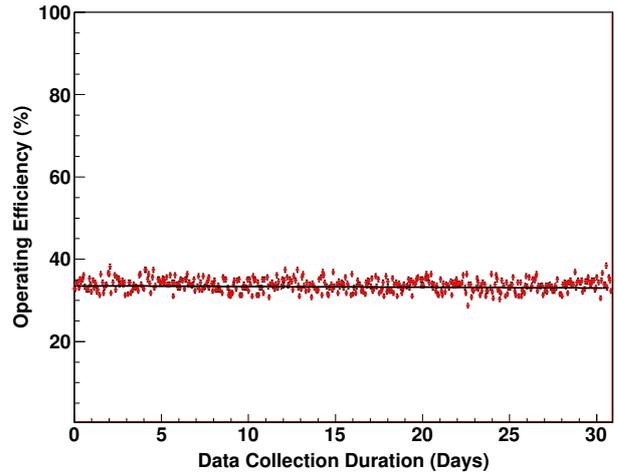}
\caption{The operating efficiency of the prototype system over a month-long data collection period.  This efficiency is defined as the percentage of events within the 3-fold trigger which contain at least one above-pedestal QDC signal in each of the other detection layers.  Each measurement was taken after 1000 registered triggers and was shown to be stable around 34\% over the duration of the study.  A first-order polynomial fit to the data is shown in the bold line.  This efficiency value translates to a mean efficiency of 80\% of the maximum achievable efficiency for each detection layer.}
\label{fig:stability}
\end{figure}

The detected efficiencies varied between 70\% and 86\% across the five layers with a stable mean operating efficiency of 34\% shown in Figure~\ref{fig:stability}.  This value is defined as the percentage of events within the 3-fold trigger which contained at least one QDC signal above pedestal in each of the other detection layers and indicated a mean detection efficiency of 80\% per layer.  The deficiencies observed relate to fibres which were damaged during the construction or assembly processes and/or dead or noisy channels.  These channels included those QDC channels connected at the extremes of the ribbon cables whose pedestals exhibited a characteristically-broader distribution. Here, the 3$\sigma$ upper threshold imposed could restrict the detection of muons which deposited low signals within the fibre.  This effect was observed in Figure~\ref{fig:sigmas} which compares the $\sigma$ values for each channel before and after optimisation of the setup.  A further source of signal loss could have been a potentially weak signal originating from the muon interacting with a small volume of scintillating material close to the edge of the fibre.  This 80\% efficiency could therefore translate to the detection of a muon signal only when it had interacted with the central 80\% (by diameter \ie 1.55\,mm) of the active circular cross-section of the fibre.  However, this assumption cannot be verified with current experimental data.

Further studies were performed by changing the innermost detection layer which was included in the trigger scheme, thus allowing the efficiency of the layer regularly used for triggering to be determined.  This allowed the efficiency of this layer to be determined, and it was found to be consistent with the range identified previously.  Position-sensitive studies, similar to those shown in Figure~\ref{fig:fibretest}, were performed using small external scintillator paddles on the outermost detector layers which confirmed these layers' consistency with the six internal layers.

\begin{figure}[t] 
\centering 
\includegraphics[width=\columnwidth,keepaspectratio]{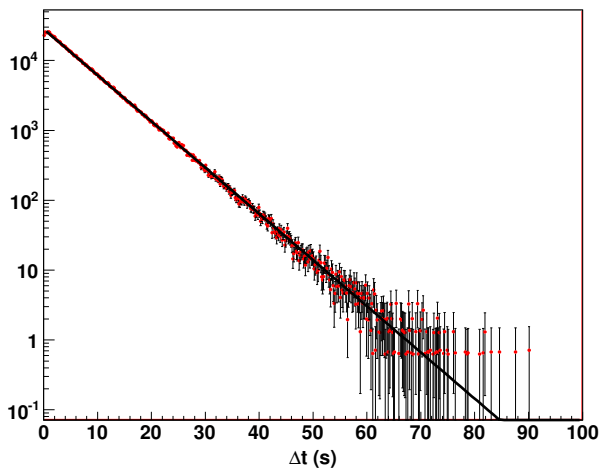}
\caption{Semi-logaritmic distribution of $\Delta$t, the time between successive triggers using the three-fold condition described in the text.  The black line through the data points represents a second-order polynomial fit to the data.  A stable rate of 0.15\,Hz was observed over the entire data collection period, corresponding to a mean $\Delta$t of approximately 6.57\,s.}
\label{fig:rates}
\end{figure}

\begin{figure}[b] 
\centering 
\includegraphics[width=0.49\columnwidth,keepaspectratio]{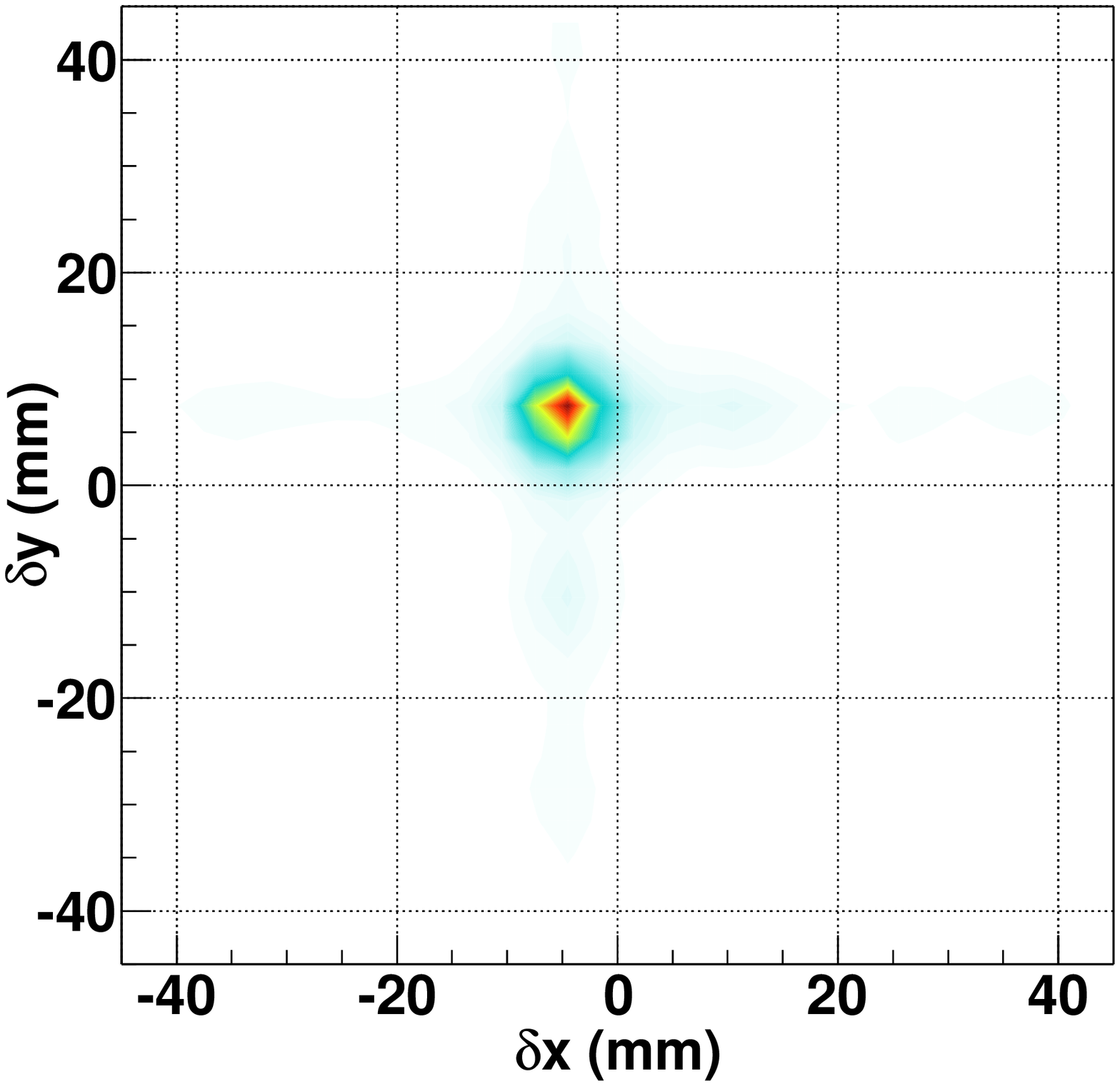}
\includegraphics[width=0.49\columnwidth,keepaspectratio]{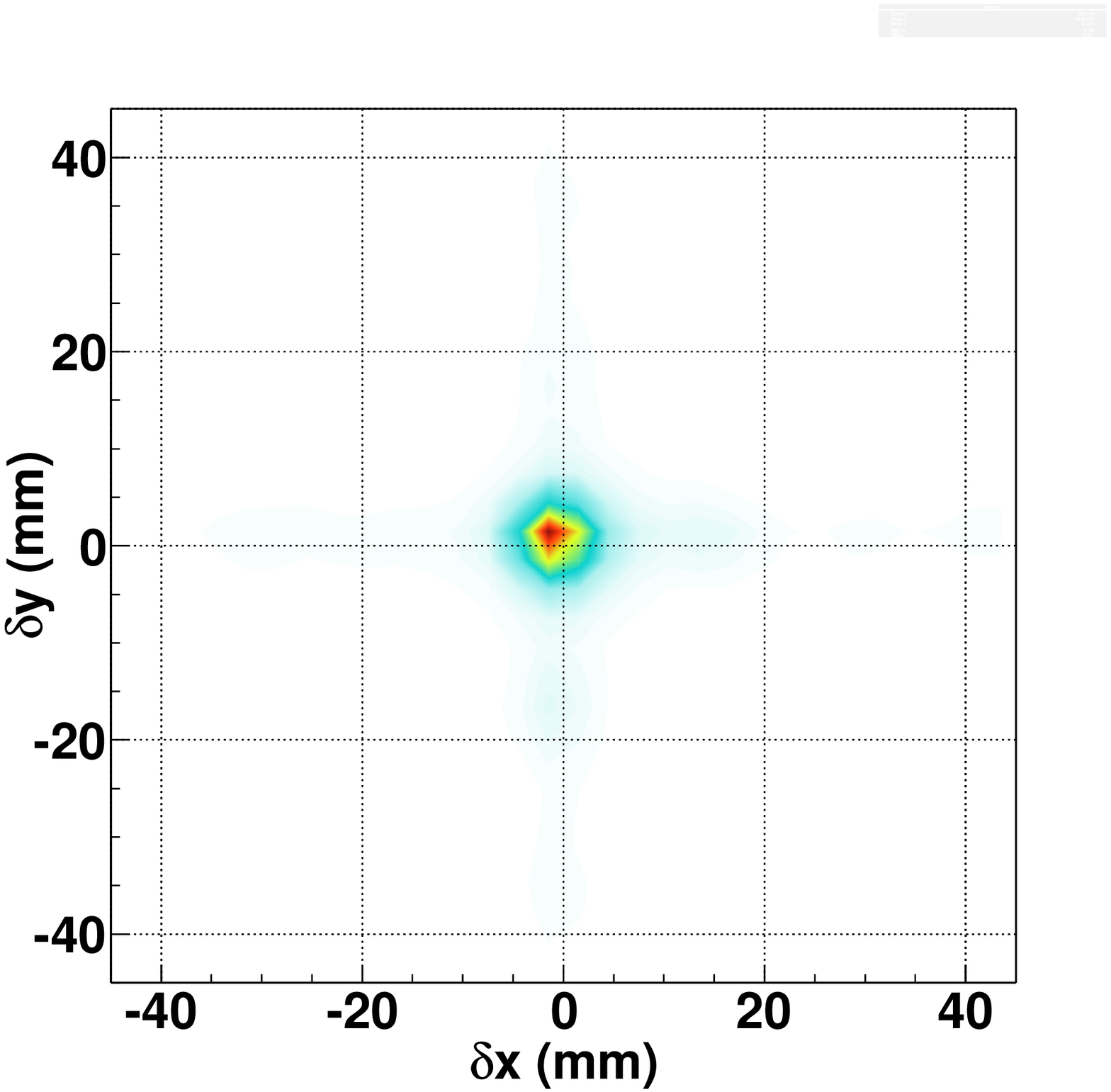}
\caption{Residual distribution from the space point on the module above the assay volume projected by the vector formed by the module pair below the assay volume prior to alignment correction (left) and post correction (right).  The slight tails on the distribution along the $\delta$x and $\delta$y axes arise from the small degree of misidentification of the hit fibre from the demultiplexing algorithm and/or from selection of a noise signal in one or more of the PMTs.}
\label{fig:Alignment}
\end{figure}

\subsection{Stability \& Detection Rates}
Throughout data collection, the prototype MT system exhibited high levels of stability which was observed in various facets of the data including the detection rates and efficiencies shown in Figure~\ref{fig:stability} for a one month period.   A steady 3-fold trigger rate of 0.15\,Hz was recorded which equated to approximately 6000 candidate tracking events (\ie with a signal in every layer) per day.  The interaction rate of cosmic muons with matter follows a Poisson-distribution. Thus the time of arrival between two events $\Delta$t exhibits an exponential decay. From data collected, the semi-logarithmic distribution of $\Delta$t is shown in Figure~\ref{fig:rates} with a mean $\Delta$t of 6.57\,s.  Nevertheless, the time between two events might exceed several tens of seconds.

\subsection{Alignment}\label{sub:Alignment}
Data were collected in 2012 with the assay volume empty of material to establish the extent of any misalignment between detector modules.   Despite the precision fabrication of the aluminium profile frame and the individual detector modules, minor misalignments were observed from the analysed data.  These were determined via an iterative process of constructing and projecting the track from one module pair to the four other detector layers and minimising the residuals $\delta$x and $\delta$y, defined as the difference between the measured and projected positions in each plane.  Once the misalignments in these layers were accounted for, the process was repeated using the track formed by these modules.  This process continued until the residuals in all eight layers were no greater than the fibre pitch.   The largest misalignment observed was approximately 5\,mm in both the x and y planes of one module.  This is shown before and after alignment correction for this particular module in Figure~\ref{fig:Alignment}.  The small misalignments identified revealed the internal alignment of the fibre planes within the module to be negligible in relation to the external alignment of the modules themselves.  The small fraction of misidentified fibres arise from the demultiplexing algorithm and from noise hits which manifest as tails along the $\delta$x and $\delta$y axes in Figure~\ref{fig:Alignment}.   With a small scattering angle assumption in place for data taken with material within the assay volume, placing a restriction on these tails further reduced the influence from fibre misidentification.

\begin{figure}[t]
\centering 
\includegraphics[width=\columnwidth,keepaspectratio]{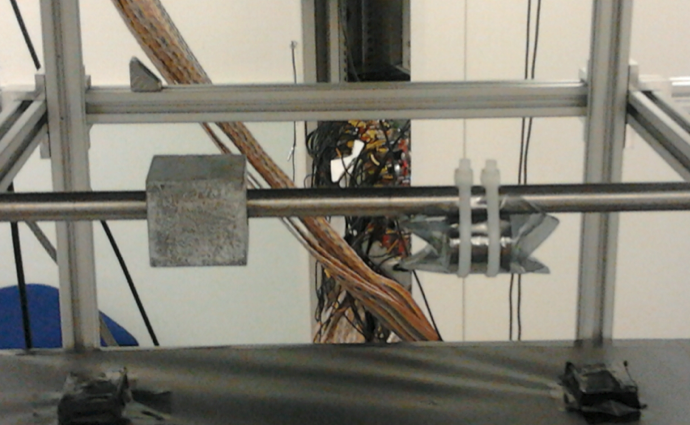}
\caption{Test configuration of materials placed in the central region of the assay volume comprising a 40\,mm cube of lead (left \ie the negative $y$-direction) surrounding a 12\,mm diameter stainless steel cylindrical rod and a 30\,mm long, 20\,mm diameter cylinder of machined uranium metal (right \ie the positive $y$-direction) suspended below.  The aluminium profile frame, ribbon cables, Tedlar\textsuperscript{\textregistered} and locating pin coverings described in the text are also observed.}
\label{fig:testsetup}
\end{figure} 
\section{Image Reconstruction of Test Objects}\label{sec:Images}
In preparation for image reconstruction studies, data collection commenced in late-2012 with a test configuration of objects placed within the assay volume.  This setup is shown in Figure~\ref{fig:testsetup}.   This consisted of a stainless-steel cylindrical bar measuring 12\,mm in diameter positioned through a 40\,mm cube of lead.  A machined cylinder of uranium metal, 20\,mm in diameter and 30\,mm in length, was suspended beneath the bar.  This bar was fixed to the aluminium profile frame along the y-direction, centred on z\,=\,0\,mm and x\,=\,$-$5\,mm in the coordinate frame used throughout this work. 

\begin{figure*}[t] 
\centering 
\includegraphics[width=1.0\columnwidth,keepaspectratio]{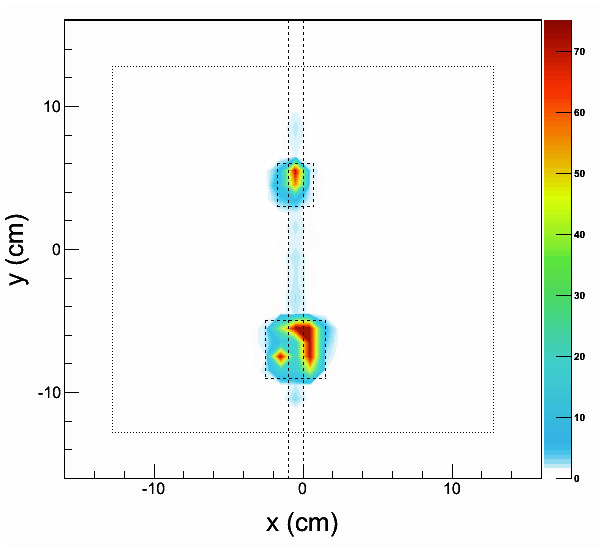}
\includegraphics[width=1.0\columnwidth,keepaspectratio]{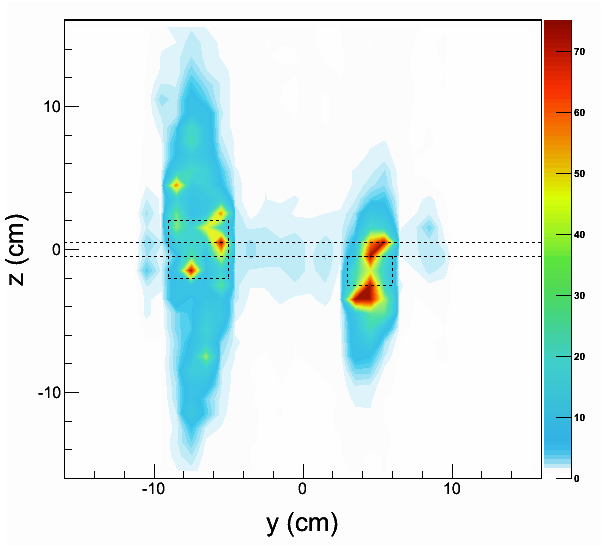}
\caption{Images reconstructed from several weeks of exposure to cosmic-ray muons.  Shown are 10\,mm slices in the xy-plane \ie parallel to the detector modules (left) and in the $yz$-plane.  This latter image should be directly compared to the test configuration shown in Figure~\ref{fig:testsetup}.   The dashed lines outline the location and dimensions of the test objects in the assay volume.  The active area of the detector system is also indicated by the fine-dotted square on the left image.  The cube of lead, the smaller cylinder of uranium and stainless-steel bar are all clearly visible within the air matrix ($\lambda <$ 1\,mrad$^{2}$\,cm$^{-1}$).  Here, the colour scale denotes the most-likely $\lambda$ value in each voxel reconstructed by the MLEM algorithm.}
\label{fig:image}
\end{figure*}

An image reconstruction algorithm based on the probabilistic Maximum Likelihood Expectation Maximisation (MLEM) method introduced in Ref.~\cite{Schultz07} by Schultz \etal was further developed for this application.  The assay volume was chosen as a cube of dimension 300\,mm in the central region of the detector assembly.  Prior to the imaging analysis, this volume was divided into small volume elements (alternatively, voxels).  Voxel dimensions are predetermined by the analyser and are at the heart of the trade-off between image resolution and measurement time; smaller voxels provide potentially greater definition up to the resolution of the detector system but require correspondingly greater measurement times.  Cubic voxels, 10\,mm in dimension, were used throughout this work as simulation showed that this size provided the best compromise between resolution and data collection time. 

For each muon, the incoming and scattered vectors were first projected to their Point of Closest Approach (PoCA).  The MLEM method then requires the calculation of a normalised scattering probability in every voxel that the muon was determined to have passed through.  After many muons, the most likely scattering density in each voxel was determined.  This was used as the imaging metric in the final results presented in this work.

\subsection{Results}\label{results}
First image reconstruction results from experimental data taken using this prototype detector system are presented in Figure~\ref{fig:image}.   Shown are two 10\,mm slices (or tomograms), one through the xy-plane (\ie\,parallel with the detection planes) which encompassed the stainless-steel bar, and the other in the yz-plane.  These images have been reconstructed from several weeks of exposure to cosmic-ray muons.  Sensitivity to atomic number $Z$ and discrimination between the $\lambda$ values of the stainless-steel bar, the two high-$Z$ material blocks, and the surrounding air is observed.  The non-uniformity of the $\lambda$ values of the reconstructed high-$Z$ objects is attributed to a combination of possible factors;  a spread in $\lambda$ values for high-density materials as a result of increased Coulomb scattering (here, the PoCA input to the MLEM method reconstructed only the average scatter) and non-uniform voxel coverage of the objects (\eg\,for the uranium object, the central voxel with a reconstructed $\lambda$ in excess of 70\,mrad$^{2}$\,cm$^{-1}$ was assumed to fully occupy the uranium, whereas the surrounding voxels occupy a combination of uranium and air which acted to dilute the reconstructed $\lambda$ value).

In the presented tomograms, the high image-resolution of this detector system in the xy-plane is observed.  Smearing and stretching of the image in the z-direction is also noted.  This is an inherent effect associated with the reconstruction of the scattering location in the principle axis of two near-parallel tracks, and is artificially exaggerated in this work by the small angular acceptance of the prototype system.   Studies which address this issue, and dedicated simulation studies using this detector geometry, are the subject of ongoing work and as such will not be described here, other than to highlight the consistency with expectation.  This effect was observed with images reconstructed from dedicated simulation studies of this test configuration of objects~\cite{SimPaper,Mahon13} which also provided excellent agreement with the image shown in Figure~\ref{fig:image}.

\section{Summary}
The design, fabrication and assembly processes undertaken to develop a modular tracker system for use in the cosmic-ray muon tomography of legacy nuclear waste containers has been presented.  The system comprised four modules with orthogonal layers of 2\,mm-pitch round scintillating fibres.  These were supported within a low-$Z$ structure fabricated using thin sheets of \roha and aluminium, and covered with a lightproof \Tedlar film and black nylon tubing.  The fibre signals were read out to QDC modules and Hamamatsu H8500 MAPMTs with two fibres coupled to a single pixel, and one PMT (\ie 128 fibres) per detection layer.  The modules were supported in an adjustable vertical support stand made from aluminium profile.  Performance studies revealed a high level of stability, both structurally and in relation to the data collected over prolonged periods of time.  High muon detection efficiencies of up to 86\% per layer were recorded with low levels (less than 5\%) of misidentification and misalignment (less than 5\,mm in the worst case prior to correction in software).  First images reconstructed from data collected with a test configuration of materials within the assay volume verified the high-$Z$ detection capabilities of this system and revealed promising low, medium and high-$Z$ discrimination which will be investigated further in future work.  All these studies will directly influence the design and construction of a large-scale detector system which will assay an industrial waste container in preparation for the industrial deployment of this technology.

\section*{Acknowledgements}
The authors gratefully acknowledge Sellafield Ltd., on behalf of the UK Nuclear Decommissioning Authority, for their continued funding of this project.


\end{document}